\documentclass[11pt]{article}

\usepackage{graphicx}
\usepackage{amsmath,amssymb}
\usepackage[T1]{fontenc}
\usepackage{graphics,wrapfig,times}
\usepackage{graphicx}
\usepackage{mathrsfs}

 \makeatletter
 \newcommand{\pback}[1]{{
   \let\@rrow=\leftarrowfill
   \mathchoice{\AIN@stemPullBack{#1}{\@rrow}}{\AIN@stemPullBack{#1}{\@rrow}}
     {\AIN@indxPullBack{#1}{\@rrow}}{\AIN@indxPullBack{#1}{\@rrow}}}
   \vphantom{#1}}

 \newcommand{\AIN@stemPullBack}[2]{
   \vtop{\mathsurround=0pt
   \ialign{##\crcr$\textstyle{#1}\strut$\crcr
     \noalign{\kern-0.4ex\nointerlineskip}{\tiny#2}\crcr}}}

 \newcommand{\AIN@indxPullBack}[2]{
   \vtop{\mathsurround=0pt
   \ialign{##\crcr\hfil$\scriptstyle{#1}$\hfil\crcr
     \noalign{\kern+0.4ex\nointerlineskip}{\tiny#2}\crcr}}}

\newcommand{\nn}{\nonumber}

\def\eth{\text{\dh}}
\def\thorn{\text{\th}}

\def\.{\cdot}

\def\be{\begin{equation}}
\def\ee{\end{equation}}
\def\bea{\begin{eqnarray}}
\def\eea{\end{eqnarray}}
\def\ba{\begin{array}}
\def\ea{\end{array}}

\newcommand{\eqhat}{\mathrel{\widehat\mathalpha{=}}}
\def\={\eqhat}

\begin{document}

\begin{center}
{\huge Horizon news function and quasi-local energy-momentum flux
near black hole}

\vspace{1cm}

{\large Yu-Huei Wu}\\

\vspace{0.5cm}

{\normalsize {\em School of Math}}
\\ {\normalsize {\em University of Southampton}}
\\ {\normalsize {\em SO17 1BJ , UK}}
\\ {\normalsize {\em E-mail:} \texttt{yw1@soton.ac.uk, yuhueiwu@hotmail.com}}
\end{center}

\begin{abstract}
From the 'quasi-local' definition of horizons, e.g. isolated
horizon and dynamical horizon, the consequence quasi-local
energy-momentum near horizons can be observed by using the idea of
frame alignment. In particular, we find the horizon news function
from the asymptotic expansion near horizons and use this to
describe the gravitational flux and change of mass of a black
hole.

\end{abstract}


\section{Introduction}

In this paper we apply a similar method of asymptotic expansion
near null infinity (Newman-Unti \cite{NU}) to the geometry near a
black hole to study gravitational radiation. The definition of
'quasi-local' horizons (isolated horizon and dynamical horizon)
are given by Ashtekar and Krishnan \cite{Ashtekar02, Ashtekar03}.
The covariant expression of quasi-local energy expression is given
by Nester-Witten two form \cite{Dougan&Mason} which refers to an
implicit reference frame when observing the quasi-local
quantities. What is a good reference frame near a strong
gravitating field such as a black hole? To tackle this problem, we
use the concept and idea of frame alignment to find the
asymptotically constant spinors which is done by Bramson
\cite{Bramson75a} for null infinity and try to move this work on
the horizon in Section \ref{constant}. We then be able to use the
spin frame $Z_A{^{\underline{A}}} = (\lambda_A, \mu_A)$ to give a
definition of the energy-momentum for a rotating dynamical horizon
in Section \ref{qRDH}. Particularly, we find the horizon news
function dominates the gravitational radiation of a black hole in
Section \ref{fluxRotDH}.

\section{Constant spinors near quasi-local horizons}\label{constant}

We find the compatible conditions for asymptotically constant
spinor $\lambda_A$ mainly include the Dougan-Mason's holomorphic
conditions and a time-related condition. Different types of
quasi-local horizons require different conditions \cite{Wu06}. For
example, the compatible conditions of a rotating dynamical horizon
are
\bea \thorn_0 \lambda^0_0 &=& 0, \Rightarrow \dot \lambda^0_0 -\epsilon_0\lambda^0_0 =0 \label{TCDdh}\\
\eth_0 \lambda^0_0 + \sigma_0\lambda_1^0&=&0, \;\;\; \eth_0 \lambda^0_1 - \mu_0 \lambda^0_0= 0, \label{DM}\\
\thorn_0 \lambda^0_1&=& - \bar\eth_0\lambda^0_0\label{extradh}
 \eea
where include  a time-related condition (\ref{TCDdh}) and
Dougan-Mason's holomorphic conditions (\ref{DM}) and an extra
condition (\ref{extradh}). Next, we will focus on the issues of a
rotating dynamical horizon.

\section{Quasi-local energy-momentum of a rotating dynamical
horizon}\label{qRDH}


By using the compatible constant spinor conditions for a rotating
dynamical horizon, i.e., the Dougan-Mason's holomorphic conditions
(\ref{DM}) in Section \ref{constant}, we can get the quasi-local
momentum integral near a rotating dynamical horizon based on
Nester-Witten two form is
\be\ba{lll} I(r') &=&  \frac{1}{4\pi} \oint_S (\rho' \lambda_0
\bar\lambda_{0'}+ \rho\lambda_1 \bar\lambda_{1'})
d S\\
&=&  \frac{1}{4\pi} \oint_S [-\mu_0 \lambda_0^0 \bar\lambda^0_{0'}
+ O(r')]d S_\Delta \label{one}\ea\ee
where we use $\rho=O (r')$, $\mu= \mu_0 + O(r')$, $d \;S_\Delta =
\lim_{r' \to 0} d S_{r'}$ and $dS_\Delta(v)$. By using the result
of the asymptotic expansion (NR7) \cite{Wu06}, we get the
quasi-local momentum integral on a rotating dynamical horizon
\be\ba{lll} I(r_\Delta)&=&- \frac{1}{4\pi} \oint
\frac{1}{2\epsilon_0}[\Psi^0_2 -\dot\mu_0 + \dot r_\Delta (\mu_0^2
+ \sigma'{_0}\bar\sigma'{_0}) + \eth_0\pi_0 + \pi_0\bar\pi_0-
\sigma_0\sigma'{_0}] \lambda_0^0 \bar\lambda^0_{0'} d S_\Delta.
\label{two} \ea\ee
%

\section{Energy-momentum flux near a rotating dynamical horizon}
\label{fluxRotDH}


Now, in order to make our calculation easier and use the
approximate Kerr in Bondi coordinate and the slow rotating
Kerr-Vaidya as our basis. We make a coordinate choice $r_\Delta(v)
= - \frac{1}{\mu_0}$ and make the assumption that
$\sigma'{_0}=\kappa_0=0 $
to simplify our calculation of flux formula.
%
%
Here we will also need the time related condition (\ref{TCDdh}) of
constant spinor of dynamical horizon in Section \ref{constant} and
re-scale it. Therefore $\dot{\lambda}^0_0=0$. We note that this
rescaling of spin frame is to chose a 'permissible' time so that
the energy flux can only depend on the news function. It's very
tedious but straightforward to calculate the flux expression. It
largely depends on the non-radial NP equations and the second
order NP coefficients. We substitute these equations back into the
energy-momentum flux formula to simplify our flux expression (See
\cite{Wu06} for the detail).

Apply time derivative on (\ref{one}), we get
\bea \dot I(r_\Delta) = \frac{1}{4\pi}\oint \dot \mu_0
\lambda_{0}^0 \bar\lambda_{0'}^0 d S_\Delta. \label{dotmu} \eea
Here we can see the fact that $\dot\mu_0$ is related with the mass
loss of gain, hence it is the \textit{news function}. Integrate
the above equation with respect to $v$ and use $\dot \mu_0 = \dot
r_\Delta {^{(2)}} R/2$, we then have
\bea d I(r_\Delta) =\frac{1}{8\pi} \int \;^{(2)}R \lambda_{0'}^0
\bar\lambda_0^0 d S_\Delta d r_\Delta. \eea
Ashtekar's total flux formula\cite{Ashtekar03} is
 \bea F_{matter} + F_{grav} = \frac{1}{16\pi}\int_{\Delta H}  \,^{(2)}
 R N
d^3 V \label{Ash-i}\eea
where $F_{matter} + F_{grav}$ is equal to flux $d I$ and $d^3 V =
d r_\Delta d S$ on horizon. Therefore, if $N=2 \lambda_{0}^0
\bar\lambda_{0'}^0$, then our flux formula from equation
(\ref{one}) is completely the same with Ashtekar-Krishnan's
formula (\ref{Ash-i}).

 Apply time derivative on (\ref{two}) 
%
%
and integrate the equation with respect to $v$, we have
\bea d I(r_\Delta) &=&-  \frac{1}{4\pi}\int\{ \frac{\mu_0}{2
\epsilon_0 \dot r_\Delta}[ \sigma_0\bar\sigma_0 + 4 \dot r_\Delta
\pi_0\bar\pi_0 + \Phi_{00}^0 ]\nn\\
 &+& \frac{2}{\epsilon_0\dot
r_\Delta} [\sigma_0\pi_0^2 + \bar\sigma_0\bar\pi_0^2]\}
\lambda_{0}^0 \bar\lambda_{0'}^0 d S_\Delta d r_\Delta.
\label{ch6-dI}\eea
where $d v = \frac{d r_\Delta}{\dot r_\Delta}$. The total flux of
Ashtekar-Krishnan \cite{Ashtekar03} in terms of NP in our gauge
\cite{Wu06} is
\bea F_{\textrm{total}} = \frac{1}{4\pi} \int [|\sigma|^2
+|\pi|^2+ \Phi_{00}] N d^3 V. \eea
In order to compare with Ashtekar's expression, therefore, if we
choose $N=2 \lambda_{0}^0 \bar\lambda_{0'}^0$ and
$-\frac{\mu_0}{2\epsilon_0 \dot r_\Delta}=2$, then (\ref{ch6-dI})
becomes
\bea d I(r_\Delta) &=& \frac{1}{4\pi}\int\{[ \sigma_0\bar\sigma_0
+  4 \dot r_\Delta \pi_0\bar\pi_0 + \Phi_{00}^0 ]  +
\frac{4}{\mu_0} [\sigma_0\pi_0^2 + \bar\sigma_0\bar\pi_0^2]\} N d
S_\Delta d r_\Delta.   \label{DHflux} \eea
However, here we have an extra term which is the coupling of the
shear $\sigma_0$ and $\pi_0$.

\section{Conclusion}

The Nester-Witten two-form with the compatible conditions of
constant spinors on 'quasi-local' horizons and together with the
results from the asymptotic expansion near 'quasi-local' horizons
give us the quasi-local energy-momentum and flux expressions on
'quasi-local' horizons. Dougan-Mason's holomorphic conditions tell
us how to gauge fix the quasi-local expression on each cross
section of 'quasi-local' horizons. The time related condition
tells us how the quasi-local expressions change with time along
horizon.
The news function which dominates the gravitational radiation near
a rotating dynamical horizon can be understood as the time
derivative of the expansion of the incoming null tetrad
$\dot\mu_0$ in equation (\ref{dotmu}). The gravitational flux for
a rotating dynamical horizon is obviously positive (mass gain)
from our formula if the dominate energy condition holds. The shear
square term in equation (\ref{ch6-dI}) which is related with the
gravitational radiation near dynamical horizon in our formula
matches the results of energy flux cross event horizon by using
the perturbation method [Hawking and Hartle]. The $\pi$ square
term is related with angular momentum contribution [See Ashtekar
and Krishnan \cite{Ashtekar03}]. There's a shear and $\pi$
coupling term in my flux formula for a rotating dynamical horizon
which is an extra term in Ashtekar-Krishnan's formula.


\end{document}